# Phase Diagram and High-Temperature Superconductivity of Compressed Selenium Hydrides

Shoutao Zhang[1], Yanchao Wang[1], Jurong Zhang[1], Hanyu Liu[1], Xin Zhong[1], Hai-Feng Song[4,5], Guochun Yang[2,1,*], Lijun Zhang[3,1,*], Yanming Ma[1,*]

[1]*State Key Laboratory of Superhard Materials, Jilin University, Changchun 130012, China.*

[2]*Faculty of Chemistry, Northeast Normal University, Changchun 130024, China.*

[3]*College of Materials Science and Engineering and Key Laboratory of Automobile Materials of MOE, Jilin University, Changchun 130012, China*

[4]*LCP, Institute of Applied Physics and Computational Mathematics, Beijing 100088, China.*

[5]*Software Center for High Performance Numerical Simulation, China Academy of Engineering Physics, Beijing 100088, China.*

[*]Address correspondence to: yanggc468@nenu.edu.cn, lijun_zhang@jlu.edu.cn or mym@calypso.cn

## ABSTRACT

Recent discovery of high-temperature superconductivity ($T_c$ = 190 K) in sulfur hydrides at megabar pressures breaks the traditional belief on the $T_c$ limit of 40 K for conventional superconductors, and open up the doors in searching new high-temperature superconductors in compounds made up of light elements. Selenium is a sister and isoelectronic element of sulfur, with a larger atomic core and a weaker electronegativity. Whether selenium hydrides share similar high-temperature superconductivity remains elusive, but it is a subject of considerable interest. First-principles swarm structure predictions are performed in an effort to seek for energetically stable and metallic selenium hydrides at high pressures. We find the phase diagram of selenium hydrides is rather different from its sulfur analogy, which is indicates by the emergence of new phases and the change of relative stabilities. Three stable and metallic species with stoichiometries of $HSe_2$, HSe and $H_3Se$ are identified above ~120 GPa and they all exhibit superconductive behaviors, of which the hydrogen-rich HSe and $H_3Se$ phases show high $T_c$ in the range of 40-110 K. Our simulations established the high-temperature superconductive nature of selenium hydrides and provided useful route for experimental verification.

## Introduction

Since Onnes discovered superconductivity of mercury in 1911,[1] intensive research activities were stimulated to search for new superconductors with high critical temperatures ($T_c$). With this thrust, unconventional superconductors such as cuprates[2-4]and Fe-pnictides,[5-7] in which superconducting mechanism cannot be described by Bardeen-Cooper-Schrieffer (BCS) theory[8] have attracted major attention since they often exhibit high $T_c$ values, reaching as high as 164 K (at high pressures).[9] In spite of extensive research for several decades, the superconducting mechanism in these unconventional superconductors is still controversial, preventing them from being an optimal platform for designing the superconductors with the higher $T_c$.

On the side of conventional superconductors, the situation is rather disappointing since they all have low $T_c$ values. The best-known conventional superconductor of $MgB_2$ has the $T_c$ of 39 K.[10] As such, there is a traditional belief on the $T_c$ limit of 40 K for conventional superconductors in the field.

BCS theory gives a clear count on superconducting mechanism of conventional superconductors, making the design of high-$T_c$ superconductors possible. According to BCS theory, a necessary condition for a high-$T_c$ superconductor is that the metallic compounds shall have large electron density of states at the Fermi level, high phonon frequencies and strong electron-phonon coupling. Hydrogen is the lightest element, and therefore naturally gives rise to high phonon frequencies, and also owes the unique strong bare electron-ion interaction. Ashcroft firstly proposed that solid hydrogen once being metallic under pressure has the potential to be a high-temperature superconductor.[11] Later on, the idea on metallic superconducting hydrogen was extended into hydrogen-rich compounds,[12] where metallization pressure can be significantly lowered than that in pure hydrogen. A number of hydrogen-rich compounds were subsequently predicted to be good superconductors[13] with estimated $T_c$ reaching remarkably high values (*e.g.* 64 K for $GeH_4$ at 220 GPa,[14] 107 K for $SiH_4(H_2)_2$ at 250 GPa,[15] 235 K for $CaH_6$ at 150 GPa,[16] and 204 K for $(H_2S)_2H_2$ at 200 GPa[17]). Experimental syntheses of these potential high-temperature superconductors are excitingly ongoing, but challenging.

Sulfur dihydride ($H_2S$) has not been considered as the candidate for superconducting hydrides since it was proposed to dissociation into elemental sulfur and hydrogen before the metallization.[18,19] Only recently, first-principles swarm structure searches on high-pressure structures of $H_2S$ was conducted and $H_2S$ was excluded from the elemental dissociation, making the prediction of its superconductivity with a high $T_c$ ~80 K at 160 GPa possible.[20] Shortly after this report, breakthrough electrical measurement observed high-temperature superconductivity in compressed $H_2S$ with an unprecedentedly high $T_c$ up to 190 K at megabar pressures.[21] The decrease of $T_c$ with magnetic field, and the strong isotope shift of $T_c$ suggest that $H_2S$ is a conventional superconductor. There exist two observed superconducting states: the sample prepared at low temperature of 100-150 K has a maximal $T_c$ of 150 K at 200 GPa, while the 190 K superconductivity comes from the sample prepared at high temperature of 220-300 K, which likely associates with the dissociation of $H_2S$ into $H_3S$.[21-23] This discovery has stimulated significant interest in studying the underlying superconducting mechanism[23-26] and searching for new high-$T_c$ superconductors in other dense hydride systems.

Selenium is a sister and isoelectronic element of sulfur with a larger atomic radius and a weaker electronegativity. By witness of high-$T_c$ superconductivity of sulfur hydrides (*i.e.* H-S system), a natural and immediate thought is to examine whether selenium hydrides (H-Se system) are also high-temperature superconductors at high pressures. To our best knowledge, there is less report on solid phases of H-Se system, except for the existence of three temperature-dependent $H_2Se$ phases at ambient pressure.[27]

We herein explored the hitherto unknown high-pressure phase diagram of the H-Se system via the first-principles swarm-intelligence based structure search. In the moderate pressure region, different from the H-S system all the H-Se compounds are energetically unstable against the elemental decomposition including the normally expectant $H_2Se$ stoichiometry. At megabar pressures above 120 GPa, three stable species with stoichiometries of $HSe_2$, HSe and $H_3Se$ were identified. Among them HSe with the high-symmetry $P4/nmm$ structure is the most stable phase and $H_3Se$ becomes marginally stable at high pressures. They are all metallic and exhibit superconductive behaviors. The latter two H-rich phases, benefiting from fairly strong electron-phonon coupling, are predicted to owe high $T_c$ values up to 110 K. Our simulations provide a useful roadmap for discovering high-temperature superconductors in selenium hydrides.

## Results

The variable-composition structure searches are performed at a variety of H-Se stoichiometries containing up to 4 formula units per simulation cell at 0, 50, 100, 200 and 300 GPa. At 0 GPa, we find only one stable stoichiometry with respect to the elemental decomposition, $H_2Se$ (see Supplemental Fig. S1a), which is consistent with available experimental reports.[27] It is in the $P3_121$ symmetry, where the arrangement of Se atoms is nearly the same as that of the Se-I phase[28] and H atoms are accommodated on the line of two adjacent Se to form covalent H-Se bonds and hydrogen bonds simultaneously (Fig. S1c). However, with increasing pressure this phase becomes dramatically unstable (Fig. S1b). The structure search results are summarized in the convex hulls constructed with solid $H_2$ and Se as the binary variables in Fig. 1 (and Fig. S1a). In the low-pressure region up to 100 GPa, all the stoichiometries are energetically unstable against the elemental decomposition. This is in sharp contrast to the H-S system, where all the investigated stoichiometries are stable with respect to the elemental decomposition.[22] With increasing pressure to 200 GPa, while most of phases still lie above the elemental decomposition line, three stoichiometries ($HSe_2$, HSe and $H_3Se$) show the tendency of being stabilized and move downward below the line. Eventually $HSe_2$ and $H_3Se$ become stable stoichiometries on the hull against any way of decomposition. At 300 GPa, the $HSe_2$, HSe and $H_3Se$ stoichiometries are all clearly located on the hull, and HSe emerges as the most stable phase over other species. The energetic instability of these phases in the low-pressure region and their being restabilized at high pressures may be attributed to the weaker covalent bond of H-Se than that of H-S (resulted from the larger size of Se), which is cumulatively strengthened by volume contraction under compression.

The Se-rich $HSe_2$ stoichiometry is the most stable phase at 200 GPa. Its lowest-energy structure (Fig. 2a) has a $C2/m$ symmetry, in which the sublattice of Se atoms is isostructural to the Se-IV phase[29] and H atoms passivate alternatively from both sides of the infinite zigzag Se chains. It is stabilized above 124 GPa (as in the inset of Fig. 1). The HSe stoichiometry is stabilized above 249 GPa as a highly symmetric PbO-type structure (space group $P4/nmm$, Fig. 2b). This compound, which is isostructural to superconductive Fe-based chalcogenides,[30] consists of the stack of two-dimensional layered edge-sharing $SeH_4$-tetrahedra networks. Within the layer both Se and H are four-fold coordinated. It is worth mentioning that in addition to the *P*4/*nmm* HSe phase, our structure search finds an energetically competitive $P2_1/c$ structure (though metastable, Fig. 2d). It consists of layered three-fold coordinated Se/H networks. As to the $H_3Se$ stoichiometry (stable above 166 GPa), the lowest-energy structure has a high symmetry of *Im-3m*, isostructural to that of $H_3S$,[17] where Se atoms occupy a body-centered cubic sublattice, each Se being six-fold coordinated by H. This structure is transformed from a molecular *R*3m phase as the result of pressure-induced hydrogen-bond symmetrization.

We then studied the electronic, phonon and electron-phonon coupling (EPC) properties for the lowest-energy structures of three stable stoichiometries. The band structures of them all exhibit metallic features in their stable pressure regions (see Fig. S3). Phonon calculations indicate their lattice dynamical stabilities evidenced by the absence of any imaginary phonon mode in the whole Brillouin zone.

For the most H-rich $H_3Se$ stoichiometry, most of conducting states across the Fermi level ($E_f$) have significant contribution from the H-*s* orbital, hybridizing strongly with the Se-*p* states, and there exists the combination of flat bands and steep bands[31] in proximity to the $E_f$ (Fig. 3a). These features, resembling those of isostructural $H_3S$,[17] are potentially favorable for strong EPC. By consideration of the heavier atomic weight of Se than that of S, one expects $H_3Se$ will have the generally lower phonon frequencies than those of $H_3S$ at the same pressure. This is indeed the case for the low-frequency Se-derived vibrations (below 15 THz) and mid-lying H-derived wagging and bending modes (between 15 and 45 THz) as shown in Fig. 3b. However, for the high-lying H-stretching vibrations (above 50 THz), one observes an opposite behavior, *i.e.* they show the higher frequencies and separate clearly from the mid-lying regime, unlike the situation of $H_3S$ in which the H-stretching vibrations mix together with the mid-lying phonons. This may be attributed to the stronger H-Se covalent bond as the result of the larger chemical precompression effect[12] induced by Se with the larger atomic radius. The calculated phonon linewidths (Fig. 3b) and EPC spectral function (Fig. 3c) indicate a similar mechanism of EPC to that of $H_3S$, where the high-frequency H-stretching modes give the notable contribution (31%) to the integral EPC parameter $\lambda$. This mechanism is different from the cases of superconducting $CaH_6$[16] and $SnH_4$[32] containing quasi-molecular H-units, where the mid-lying H-derived vibrations contribute most significantly to the EPC. For both of compounds, the EPC shows strong anisotropy along different phonon momentum vectors. Because of the hardening of H-stretching phonons, $H_3Se$ shows a slightly reduced $\lambda$ (1.04 compared with 1.33 of $H_3S$), but still falling in the range of fairly strong EPC. Different from the linearly decreased $\lambda$ with pressure in $H_3S$, the $\lambda$ of $H_3Se$ shows negligible pressure dependence (Fig. 3d). By substituting $\lambda$ into the Allen-Dynes modified McMillan equation,[33] we get a weakly pressure-dependent superconducting $T_c$ around 110 K for $H_3Se$, mildly lower than the values of 160-170 K for $H_3S$ (see also Table S3).

Turning to the HSe stoichiometry, there is moderate contribution of the H-*s* state to the conducting states, *i.e.* only several bands across the $E_f$ (*e.g.* along the A-M, M-Γ and X-Γ lines) are derived from the H-*s* orbital (Fig. 4a). The Fermi surface (Fig. 4b and Fig. S8) consists of the relatively small pockets around the M and Z points, and two expanding sheets at the larger wave-vectors. No notable Fermi nesting can be observed. Analysis of bonding feature via the electron localization function (ELF, Fig. 4c) indicates a much weaker H-Se covalent bond (with the maximum ELF magnitude of ~0.5) by comparison with that of $H_3Se$ (with the maximum ELF of ~0.9). This is further supported by the elongated bond length of HSe (1.69 Å) than that of $H_3Se$ (1.51 Å) at the same pressure (300 GPa). The weakening of the covalent H-Se bonding results in a remarkably softening of phonon spectrum (solid blue line in the phonon density of states plot of Fig. 4d) compared with the case of $H_3Se$ (black dash line). The EPC calculations (the upper panel of Fig. 4d) gives a moderately strong EPC parameter $\lambda$ of ~0.8 (see also Table S3), where the H-derived vibrations make a ~48% contribution. The smaller magnitude of $\lambda$ in HSe than that of $H_3Se$ can be rationalized by the weakened bonding strength in the covalent-bond system.[34] Meanwhile, due to the relatively low logarithmic averaged phonon frequency $\omega_{log}$ (800-900 K) as the result of phonon softening, HSe exhibits a moderate $T_c$ around 40 K.

For the Se-rich $HSe_2$ stoichiometry, as expected its electronic structure is predominated by Se rather than H. The calculated rather small $\lambda$ of 0.45 and low $T_c$ of ~5 K (at 300 GPa) is reminiscent of superconducting solid Se at high pressures.[35]

## Discussion

After the completion of our work, we were aware of the work by Flores-Livas *et al.* [36]predicting high-temperature superconductivity (with $T_c$ up to 131 K) in selenium hydrides. Our work is different from theirs in several aspects: (i) they only focus on the $H_3Se$ stoichiometry in analogy to $H_3S$, whereas we investigated the entire energy landscape of H-Se system by a more comprehensive global structure search. In addition to $H_3Se$, we identified two new energetically stable stoichiometries *i.e.*, $HSe_2$ and HSe; (ii) we got a qualitatively different picture of energetic stability for selenium hydrides with respect to the elemental decomposition from theirs. Most of their structures are highly stable relative to the elemental decomposition, but in our work the stable stoichiometries against the elemental decomposition can only emerge at quite high pressures (above 100 GPa); (iii) in contrast to their work assuming $H_3Se$ as the most stable stoichiometry, we find in fact $H_3Se$ is marginally stable, and $HSe_2$ and HSe are the most stable species at medium and high pressure region, respectively. The implication of our results is that the synthesis of $H_3Se$ in experiments may require particular kinetic control process.

The HSe phase in the *P*4/*nmm* symmetry represents another interesting high-symmetry structure in addition to the intriguing *Im*-3*m* structure discovered originally in $H_3S$.[17] It should be pointed out that this structure is completely different from the one of HS (in the low symmetry of *C*2/*m*) predicted by Errea *et al.*[37] For such a highly symmetric structure, the transition barrier between it and other isomers is usually high, which points to a strong kinetic stability with respect to variations of external conditions, thus favorable to experimental synthesis.

## Conclusion

In summary, with the aim of finding stable and metallic selenium hydrides for potential high-$T_c$ superconductors, we explore via a global-minimum structure search method the hitherto unknown energy landscape of H-Se system at high pressures. Despite of similar electronic properties of Se and S, the high-pressure phase diagram of H-Se system is distinct from its H-S analogy. The $H_2Se$ stoichiometry expected from the normal valence states of H (+1) and Se (-2) is surprisingly unstable against the elemental decomposition. Three energetically stable phases, *i.e.* $HSe_2$, HSe and $H_3Se$, are identified above 120 GPa. While the $H_3S$ stoichiometry dominates as the most stable phase in the H-S system, $H_3Se$ turns out to be marginally stable and the most stable stoichiometry is the highly symmetric HSe. All of stable phases exhibit metallic features and superconducting activities. The latter two are predicted to have a high $T_c$ of 40 K (HSe) and 110 K ($H_3Se$). Experimental attempt to synthesize these new phases and verification of their superconductivity are called for.

## Methods

The energetic stability of H-Se system is investigated by globally minimizing the potential energy surface at varied stoichiometries via an in-house developed swarm-intelligence based CALYPSO method[38, 39] in combination with *ab initio* density functional theory (DFT) total-energy calculations. Its validity in rapidly finding the stable ground-state structures has been demonstrated by its applications in various material systems ranging from elements to binary and ternary compounds.[38, 40-42] The energetic calculations are performed using the plane-wave pseudopotential method within the generalized gradient approximation through the Perdew–Burke–Ernzerhof (PBE) exchange-correlation functional[43], as implemented in the VASP code.[44]The electron-ion interaction was described by the projected-augmented-wave potentials with $1s^1$ and $4s^2 4p^4$ as valence electrons for H and Se, respectively. During the structure search, an economy set of parameters are used to calculate the relative energetics of sampled structures, following which the cutoff energy of 600 eVfor the expansion of wave-function and Monkhorst−Pack *k*-point sampling with grid spacing of $2\pi\times$ 0.03 $Å^{-1}$were chosen to ensure the enthalpy converged to better than 1 meV/atom. The validity of pseudopotentials used at high pressures is carefully examined with the full-potential linearized augmented plane-wave method through the WIEN2k package.[45] The phonon spectrum for evaluating the lattice dynamic stability and electron-phonon coupling for superconducting properties of stable phases are performed within the framework of the linear-response theory via Quantum-ESPRESSOpackage.[46] See Supplementary information for more details.

## Acknowledgements

This research was supported by the China 973 Program (2011CB808200), Natural Science Foundation of China under No. 11274136, the 2012 Changjiang Scholars Program of China,the Natural Science Foundation of Jilin Province (20150101042JC) and the Postdoctoral Science Foundation of China under grant 2013M541283. L.Z. acknowledges funding support from the Recruitment Program of Global Experts (the Thousand Young Talents Plan).

## Author contributions

Y. M., L. Z. and G. Y. conceived the idea and supervised the project. S. Z., Y. W., J. Z. and H. L. performed the calculations. All the authors contributed to analyzing the results. Y. M., L. Z., G. Y. and S. Z. wrote the paper.

## Competing financial interests

The authors declare no competing financial interests.

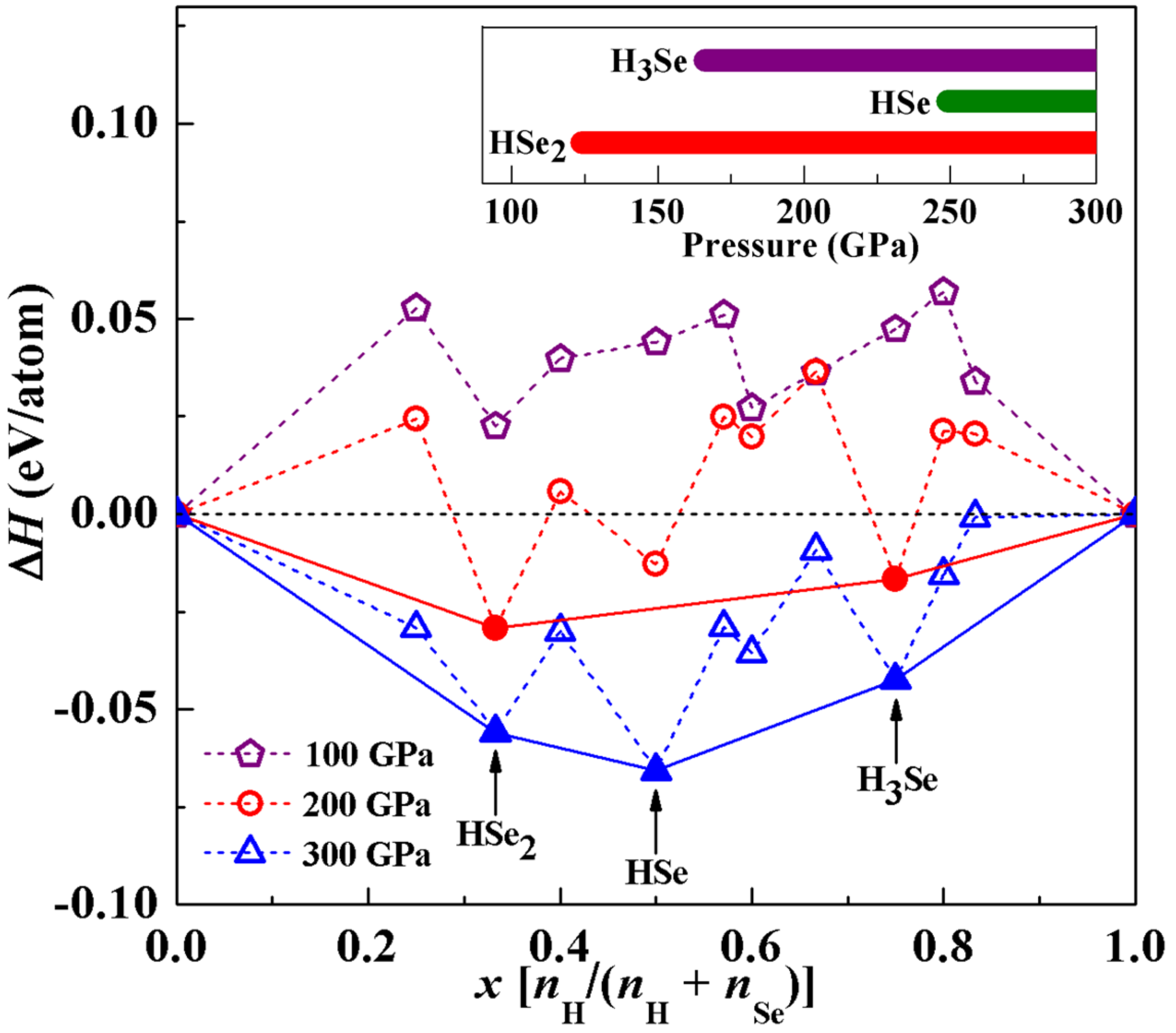

**Figure 1.** Calculated formation enthalpies ($H$ in eV/atom) of various selenium hydrides with respect to the elemental decomposition into solidified $H_2$ and Se at 100 (violet), 200 (red) and 300 (blue) GPa, respectively. At each stoichiometry, only $H$ of the lowest-energy structure is shown. The phase IV ($C2/m$) and phase VI ($Im$-$3m$) of Se,[29] the $P6_3m$ and $C2/c$ structure of solid $H_2$[47] in respective stable pressure regions are chosen for calculating $H$. The inset plot shows the pressure range in which each stable stoichiometry is stabilized.

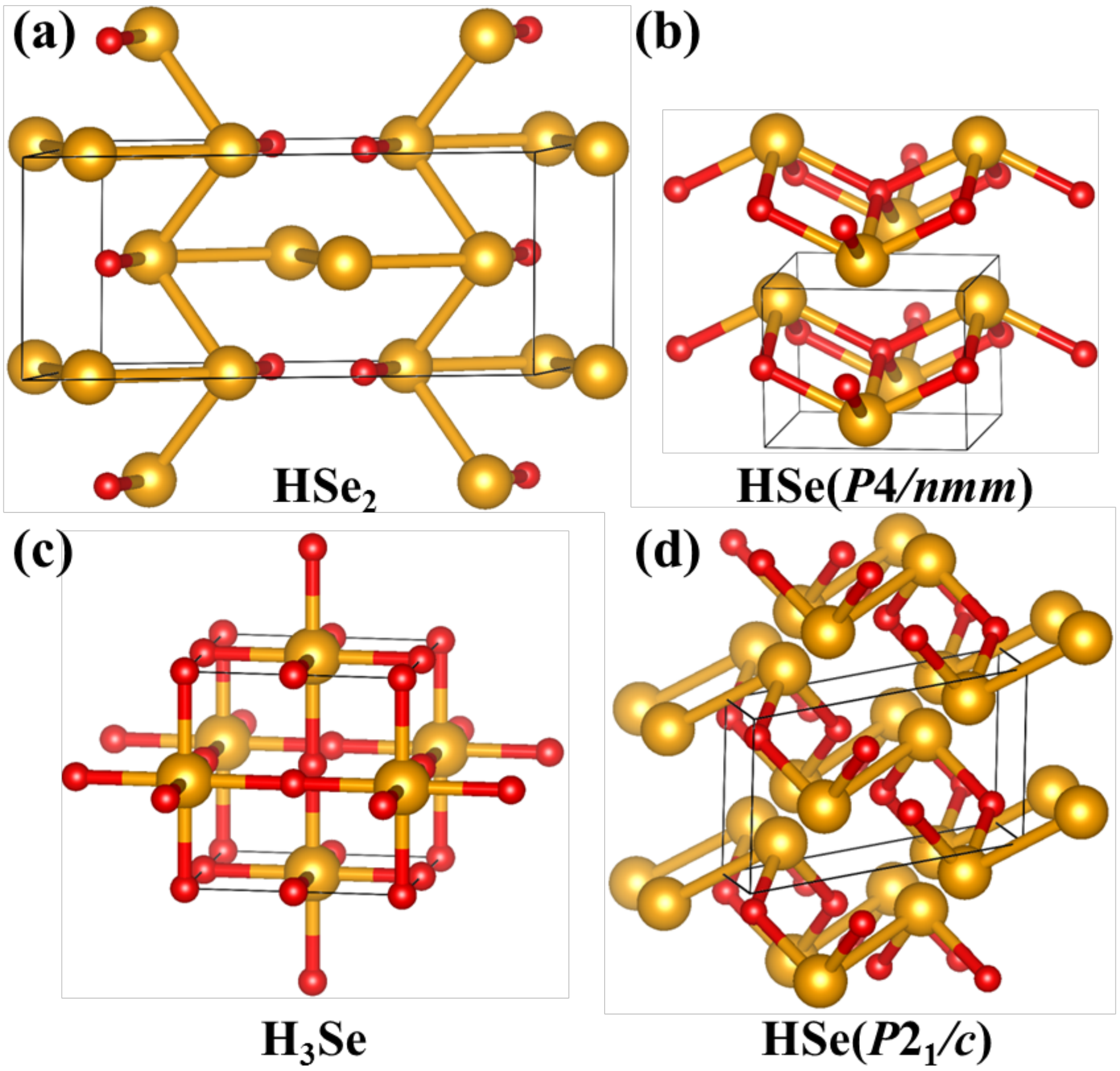

**Figure 2.** The energetically stable H-Te compounds identified by the structure search: (a) $HSe_2$ in the *C*2/*m* structure, (b) HSe in the *P*4/*nmm* structure and (c) $H_3Se$ in the *Im*-3*m* structure. For the HSe stoichiometry, the metastable $P2_1$/c structure with competitive enthalpy is shown in (d). See Supplemental Table S1 for their detailed structural information and Fig. S2 for more metastable structures and Figs S5-S7 for specific enthalpy-pressure relationship of each stoichiometry.

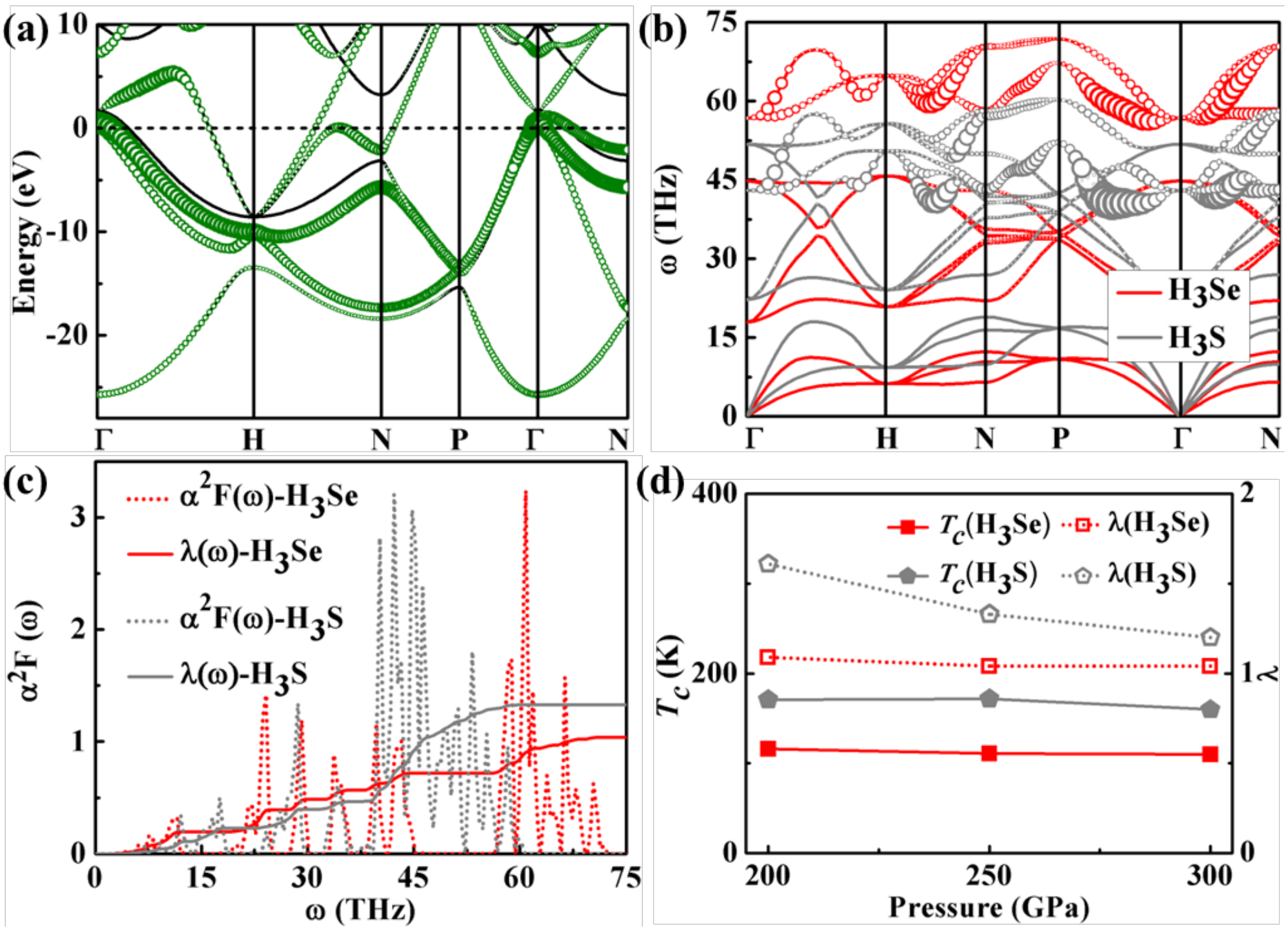

**Figure 3.** (a) Electronic band structure of $H_3Se$ in the *Im*-3*m* structure at 250 GPa. The projection onto the H-*s* orbital is depicted by the sizes of green circles. (b) Comparison of phonon spectra of $H_3Se$ (red) and $H_3S$ (gray) at 250 GPa. The phonon linewidth $\gamma_{q,j}(\omega)$ of each mode ($q$,$j$) caused by EPC is illustrated by the size of circle. (c) Eliashberg EPC spectral function $\alpha^2F(\omega)$ and EPC integration $\lambda(\omega)$ of $H_3Se$ and $H_3S$. (d) Pressure dependence of the EPC parameter $\lambda$ (right axis) and $T_c$ (left axis) for $H_3Se$ and $H_3S$. The typical value of the Coulomb pseudopotential $\mu^* = 0.1$ is used for calculating $T_c$.

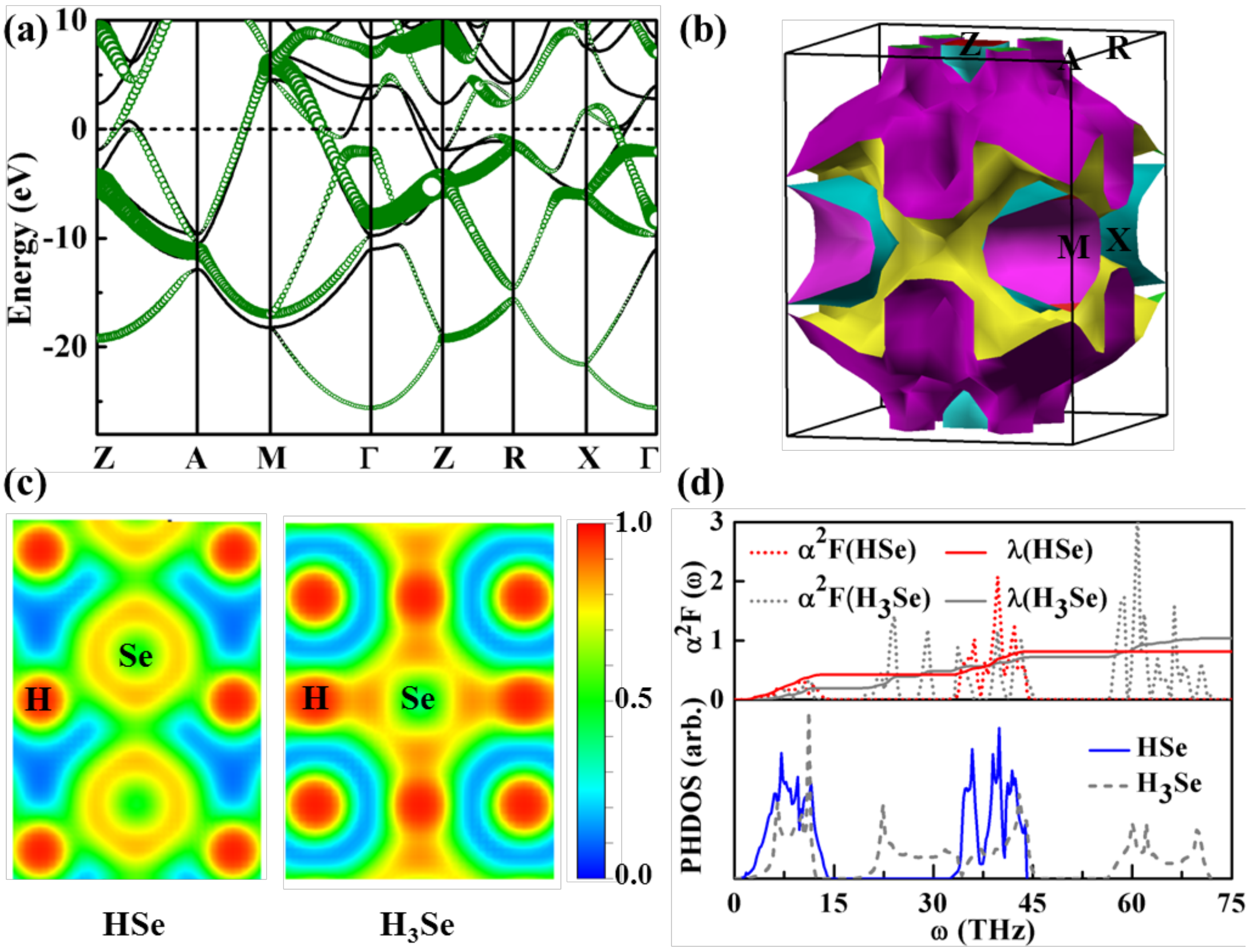

**Figure 4.** (a) Electronic band structure of HSe in the *P*4/*nmm* structureat 250 GPa. Similar to Fig. 3a, the projection onto the H-*s* orbital is indicated by green circles. (b) Fermi surface of HSe at 250 GPa. (c) Comparison of the electron localization function within the (010) plane of HSe (left) and $H_3Se$ (right) at the same pressure. (d) Phonon density of states (lower panels), Eliashberg EPC spectral function $\alpha^2F(\omega)$ and EPC integration $\lambda(\omega)$ (upper panels) of HSe and $H_3Se$.

**Supplementary Information** for the paper entitled "*Phase Diagram and High-Temperature Superconductivity of Compressed Selenium Hydrides*"

Shoutao Zhang[1], Yanchao Wang[1], Jurong Zhang[1], Hanyu Liu[1], Xin Zhong[1], Hai-Feng Song[4,5], Guochun Yang[2,1], Lijun Zhang[3,1], Yanming Ma[1]

*[1]State Key Laboratory of Superhard Materials, Jilin University, Changchun 130012, China.*
*[2]Faculty of Chemistry, Northeast Normal University, Changchun 130024, China.*
*[3]College of Materials Science and Engineering and Key Laboratory of Automobile Materials of MOE, Jilin University, Changchun 130012, China*
*[4]LCP, Institute of Applied Physics and Computational Mathematics, Beijing 100088, China.*
*[5]Software Center for High Performance Numerical Simulation, China Academy of Engineering Physics, Beijing 100088, China.*

# Supplementary Methods

Our structural prediction approach is based on a global minimization of free energy surfaces of given compounds by combining *ab initio* total-energy calculations with the particle swarm optimization (PSO) algorithm.[1,2] The structure search of each $H_xSe_y$ ($x$ = 1−5 and $y$ = 1−3) stoichiometry is performed with simulation cells containing 1−4 formula units. In the first generation, a population of structures belonging to certain space group symmetries are randomly constructed. Local optimizations of candidate structures are done by using the conjugate gradients method through the VASP code[3], with an economy set of input parameters and an enthalpy convergence threshold of $1 \times 10^{-5}$ eV per cell. Starting from the second generation, 60% structures in the previous generation with the lower enthalpies are selected to produce the structures of next generation by the PSO operators. The 40% structures in the new generation are randomly generated. A structure fingerprinting technique of bond characterization matrix is employed to evaluate each newly generated structure, so that identical structures are strictly forbidden. These procedures significantly enhance the diversity of sampled structures during the evolution, which is crucial in driving the search into the global minimum. For most of cases, the structure search for each chemical composition is converged (evidenced by no structure with the lower enthalpy emerging) after 1000 ~ 1200 structures investigated (*i.e.* in about 20 ~ 30 generations).

The energetic stabilities of different $H_xSe_y$ stoichiometries are evaluated by their formation enthalpies relative to the products of dissociation into constituent elements (*i.e.* solidified phases of $H_2$ and Se):

$$\Delta H = [h(H_xSe_y) - xh(H) - yh(Se)]/(x+y) \quad (1)$$

where $h$ represents absolute formation enthalpy. By regarding H and Se as the binary variables, with these $\Delta H$ values we can construct the convex hull at each pressure (Fig. 1in the main text). It is known that the zero-point energy plays an important role in determining the phase stabilities of the compounds containing light elements such as

H. We hence examine the effect of zero-point energy on the stability of the stoichiometries on the convex hull (Fig. S9), by using the calculated phonon spectrum with the supercell approach[6] as implemented in the Phonopy code[7].

The electron-phonon coupling calculations are carried out with the density functional perturbation (linear response) theory as implemented in the QUANTUM ESPRESSO package.[8] We employ the norm-conserving pseudopotentials with $1s^1$ and $4s^2 4p^4$ as valence electrons for H and Se. The kinetic energy cutoff for wave-function expansion is chosen as 70 Ry. To reliably calculate electron-phonon coupling in metallic systems, we need to sample dense *k*-meshes for electronic Brillouin zone integration and enough *q*-meshes for evaluating average contributions from the phonon modes. Dependent on specific structures of stable compounds, different *k*-meshes and *q*-meshes are used: 16 x 16 x 16 *k*-meshes and 4 x 4 x 4 *q*-meshes for $HSe_2$ in the *C*2*/m* structure, 18 x 18 x 24 *k*-meshes and 3 x 3 x 6 *q*-meshes for HSe with the *P*4*/nmm* structure, 24 x 24 x 24 *k*-meshes and 6 x 6 x 6 *q*-meshes $H_3Se$ with the *Im*-3*m* structure.

To examine the reliability of the adopted projected-augmented-wave (PAW) pseudopotentials for H and Se, the formation enthalpy calculations of $H_3Se$ with other types of pseudopotentials (contained in the VASP code) are performed. The results are shown in Table S0. As seen different pseudopotentials generally give consistent results, though the LDA ones turns out to overcount the magnitude of formation enthalpy. The reliabilities of the pseudopotentials at high pressures are also crosschecked with the full-potential linearized augmented plane-wave (LAPW) method as implemented in the WIEN2k code[9]. By using the two different methods, we calculate total energies of $H_3Se$ in the *Im*-3*m* structure with varying pressures, and then fit the obtained energy-volume data into the Birch-Murnaghan equation of states. Figure S0 shows the resulted fitted equation of states. We can see the results derived from two methods are almost identical, which clearly indicates the suitability of the PAW pseudopotentials for describing the energetics of Se hydrides at megabar pressures.

**Table S0.** Calculated formation enthalpy (in eV) per formula unit of $H_3Se$ with respect to solid $H_2$ and Se using different PAW pseudopotentials at 150 GPa.

| **PBE** | **H** | **H_s** | **H_h** | **H_GW** |
|---|---|---|---|---|
| **Se** | -0.0092 | -0.0193 | -0.0105 | -0.0105 |
| **Se_GW** | -0.0079 | -0.0180 | -0.0093 | -0.0093 |
| **LDA** | **H** | **H_s** | **H_h** | **H_GW** |
| **Se** | -0.1531 | -0.1538 | -0.1536 | -0.1538 |
| **Se_GW** | -0.1528 | -0.1534 | -0.1533 | -0.1534 |

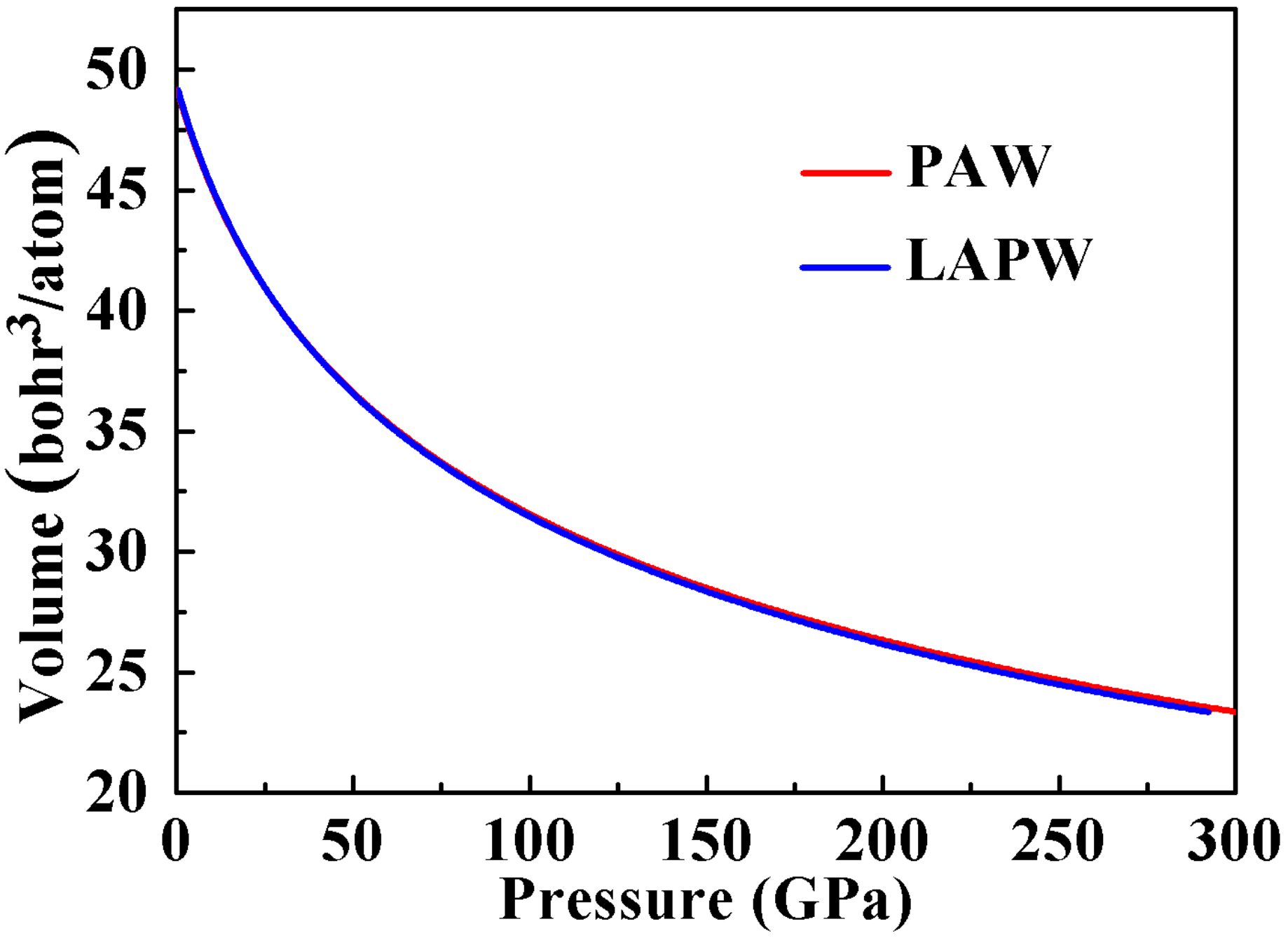

**Figure S0.** Comparison of the fitted Birch-Murnaghan equation of states for $H_3Se$ in the *Im*-3*m* structure by using the calculated results from the PAW pseudopotentials and the full-potential LAPW methods.

# Supplementary Figures

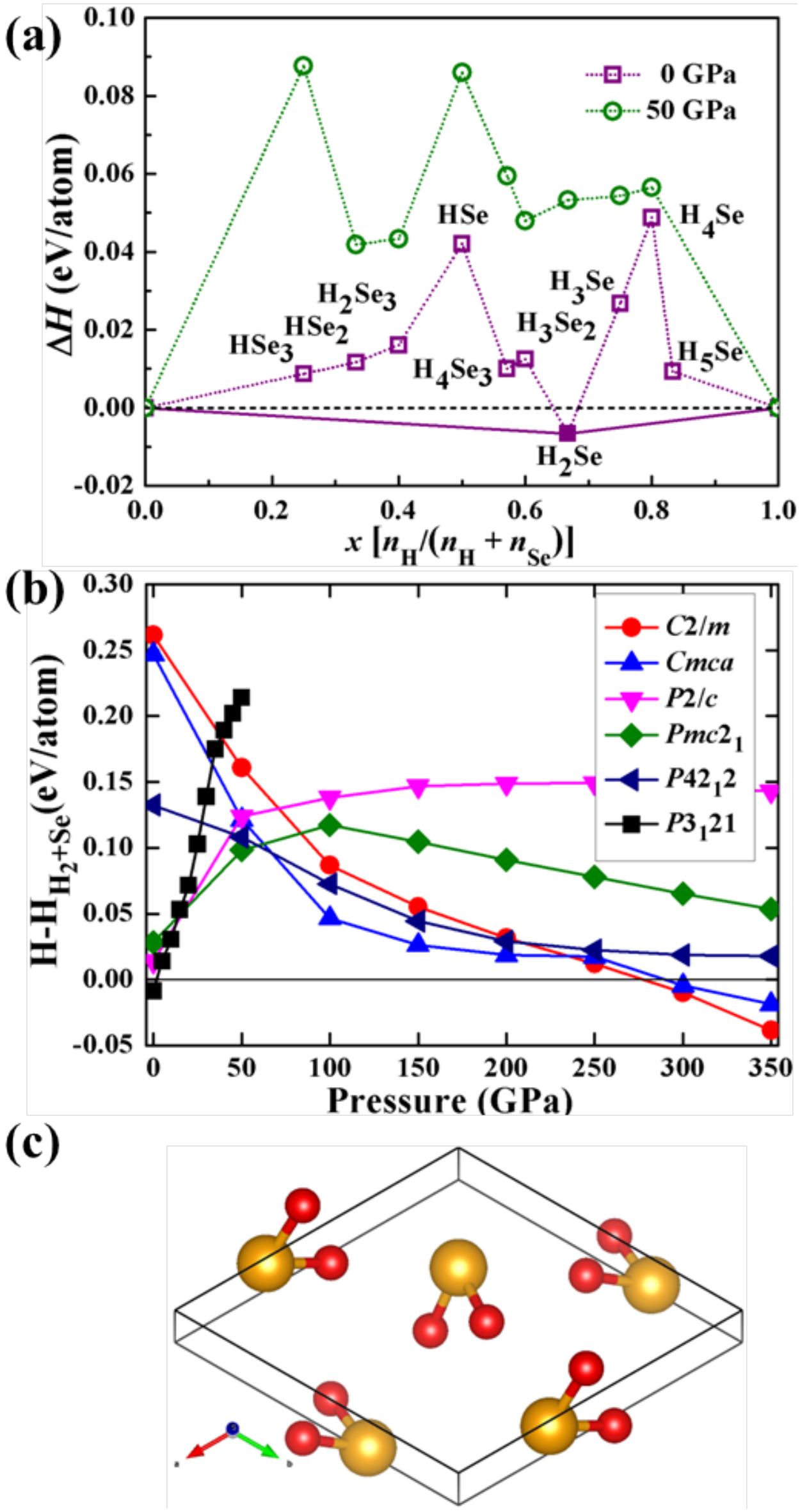

**Figure S1.** (a) Phase stabilities of various H-Se compounds at the low pressure range of 0 and 50 GPa. The formation enthalpies of H-Se compounds are relative to the enthalpies of elemental decomposition into solidified phases of $H_2$ and Se. Dashed lines connect data points, and solid lines denote the convex hull. Compounds corresponding to data points located on the convex hull are stable against decomposition. (b) Calculated enthalpies per atom of various structures for stoichiometry $H_2Se$ in the pressure range of 0–350 GPa with respect to $H_2$+Se. (c) Predicted structure of $H_2Se$ with $P3_121$ symmetry at 0 GPa.

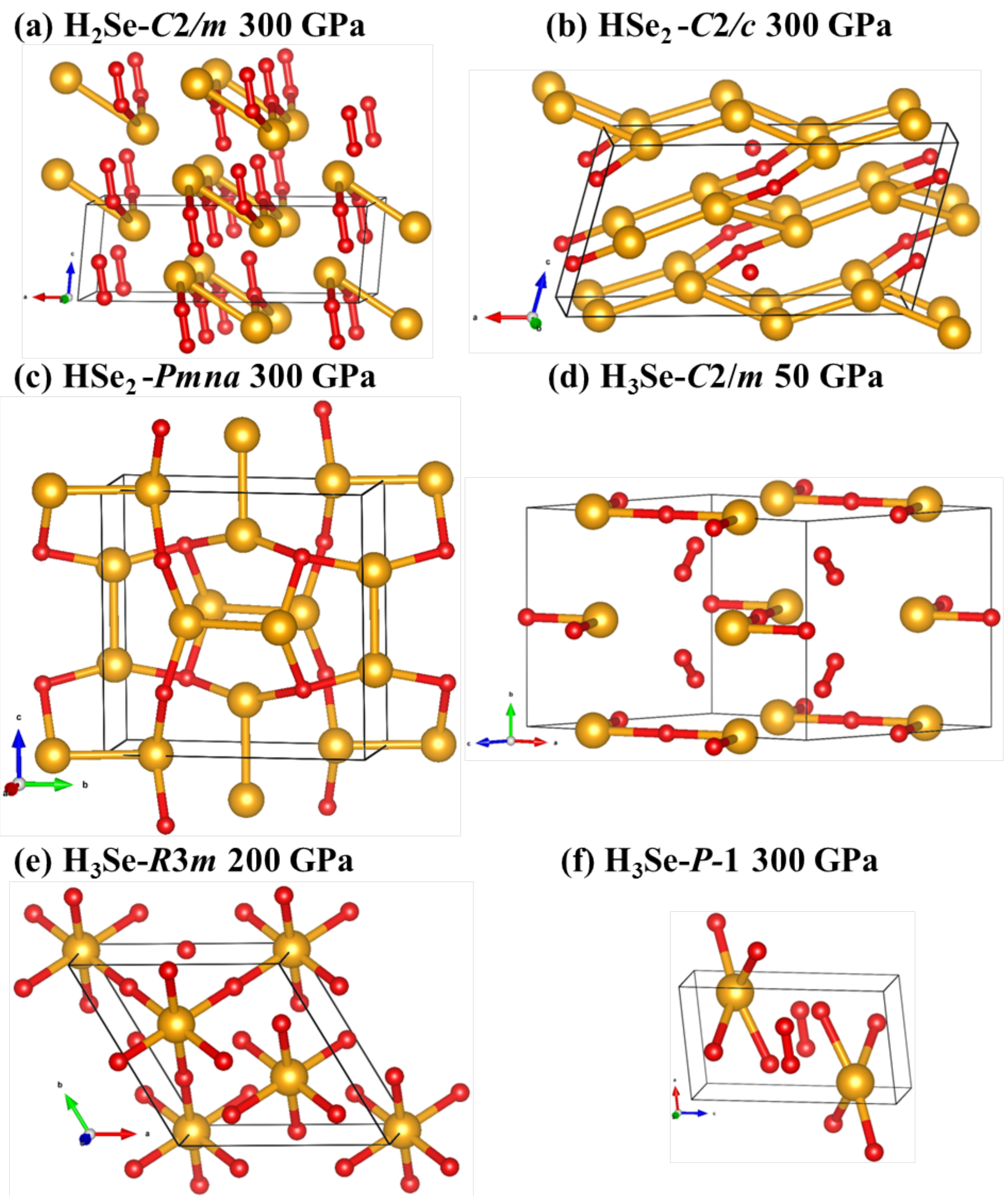

**Figure S2.** The metastable structures for stoichiometries $H_2Se$, $HSe_2$ and $H_3Se$ at selected pressures. The corresponding detailed structural information is summarized in Table S2.

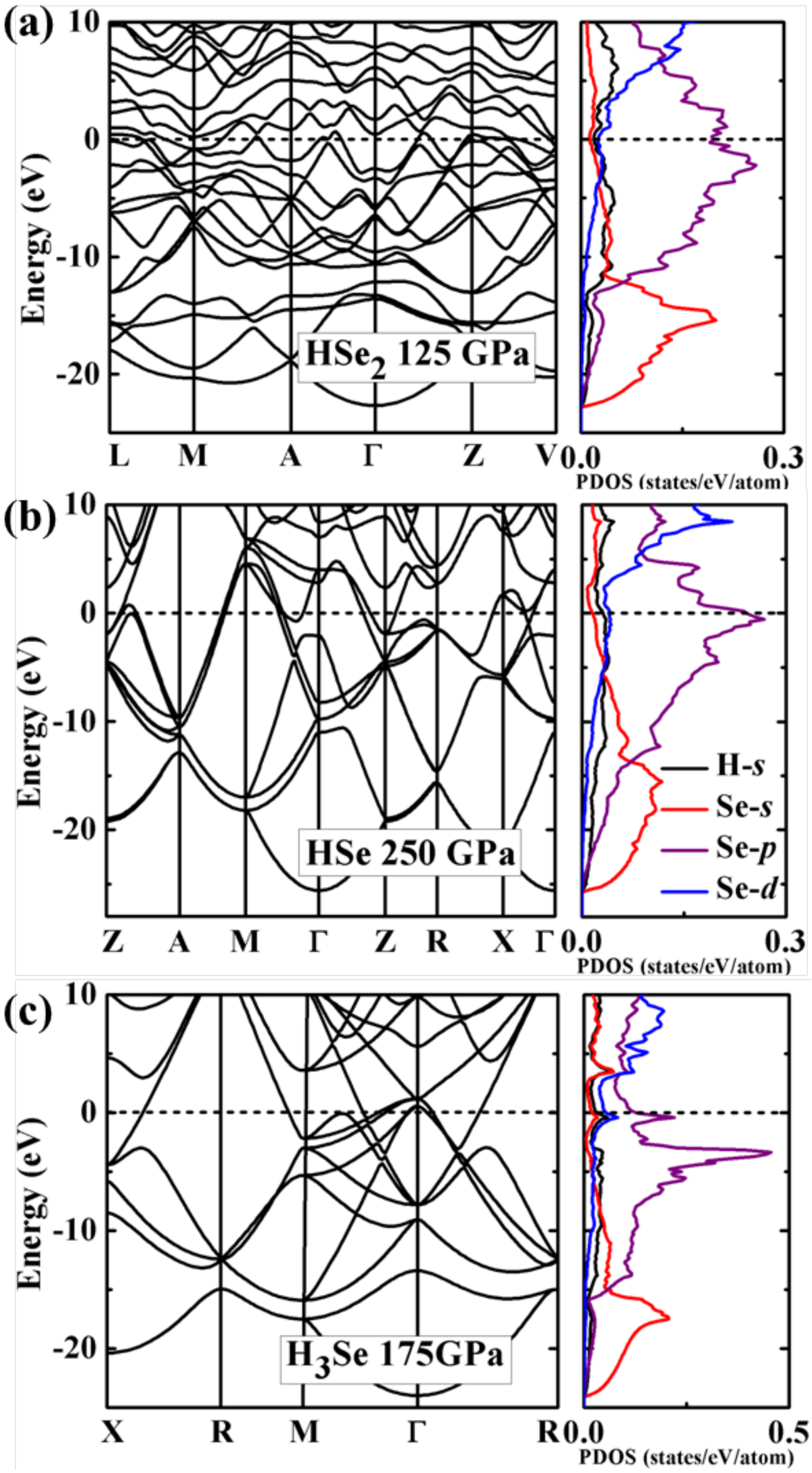

**Figure S3.** Calculated electronic band structure (left panels) and projected density of states (right panels) for the predicted H-rich compounds at the lower boundary of respective stable pressure region: (a) $HSe_2$(*C*2/*m*) at 125 GPa, (b) HSe(*P*4/*nmm*) at 250 GPa, and (c) $H_3Se$(*Im*-3*m*) at 175 GPa.

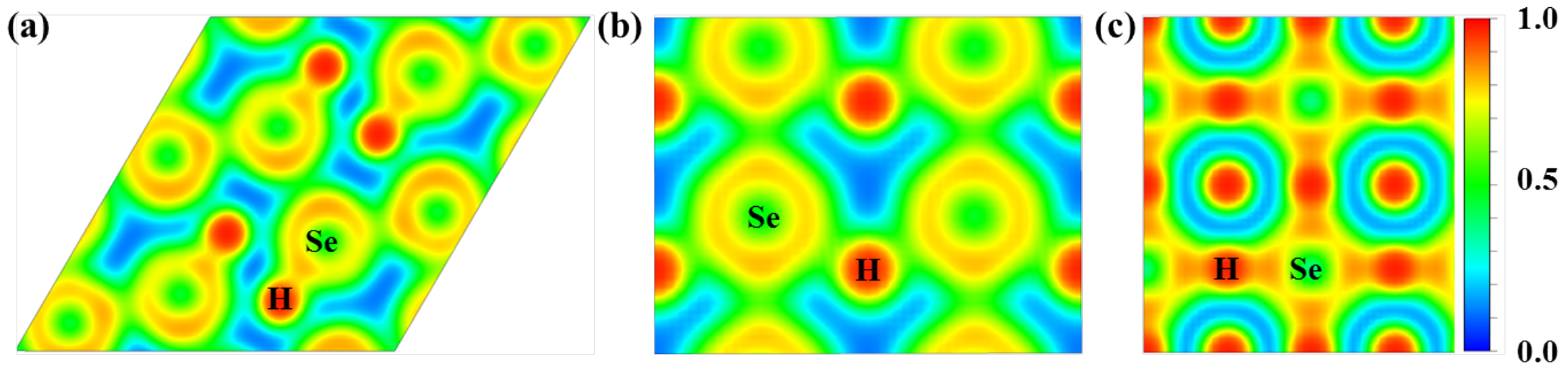

**Figure S4.** Two-dimensional plots of the electron localization function (ELF) for $HSe_2$(*C*2/*m*), HSe(*P*4/*nmm*) and $H_3Se$(*Im*-3*m*) at 300 GPa.

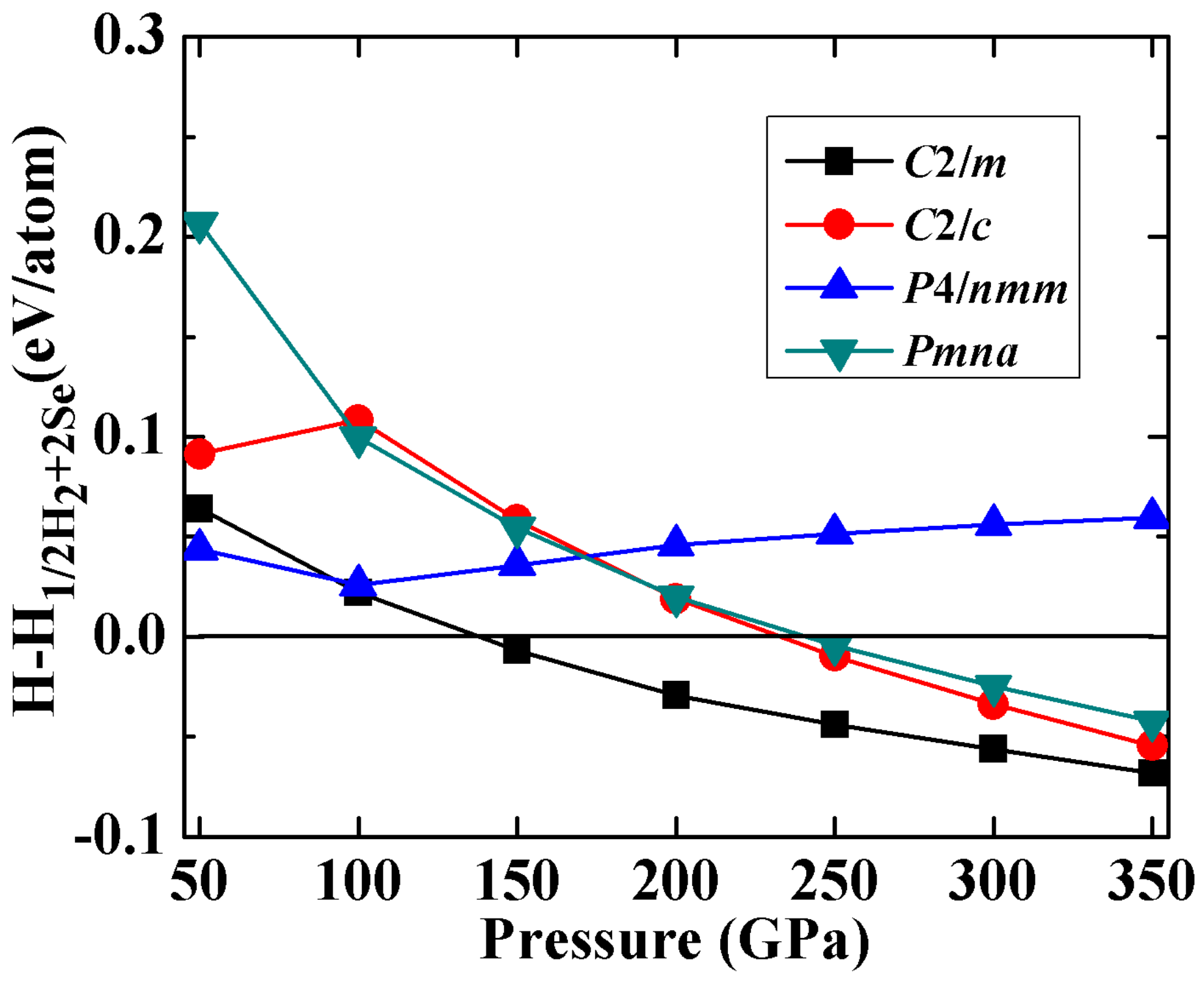

**Figure S5.** Calculated enthalpies per atom of various structures for stoichiometry $HSe_2$ in the pressure range of 50–350 GPa with respect to $1/2H_2$+2Se.

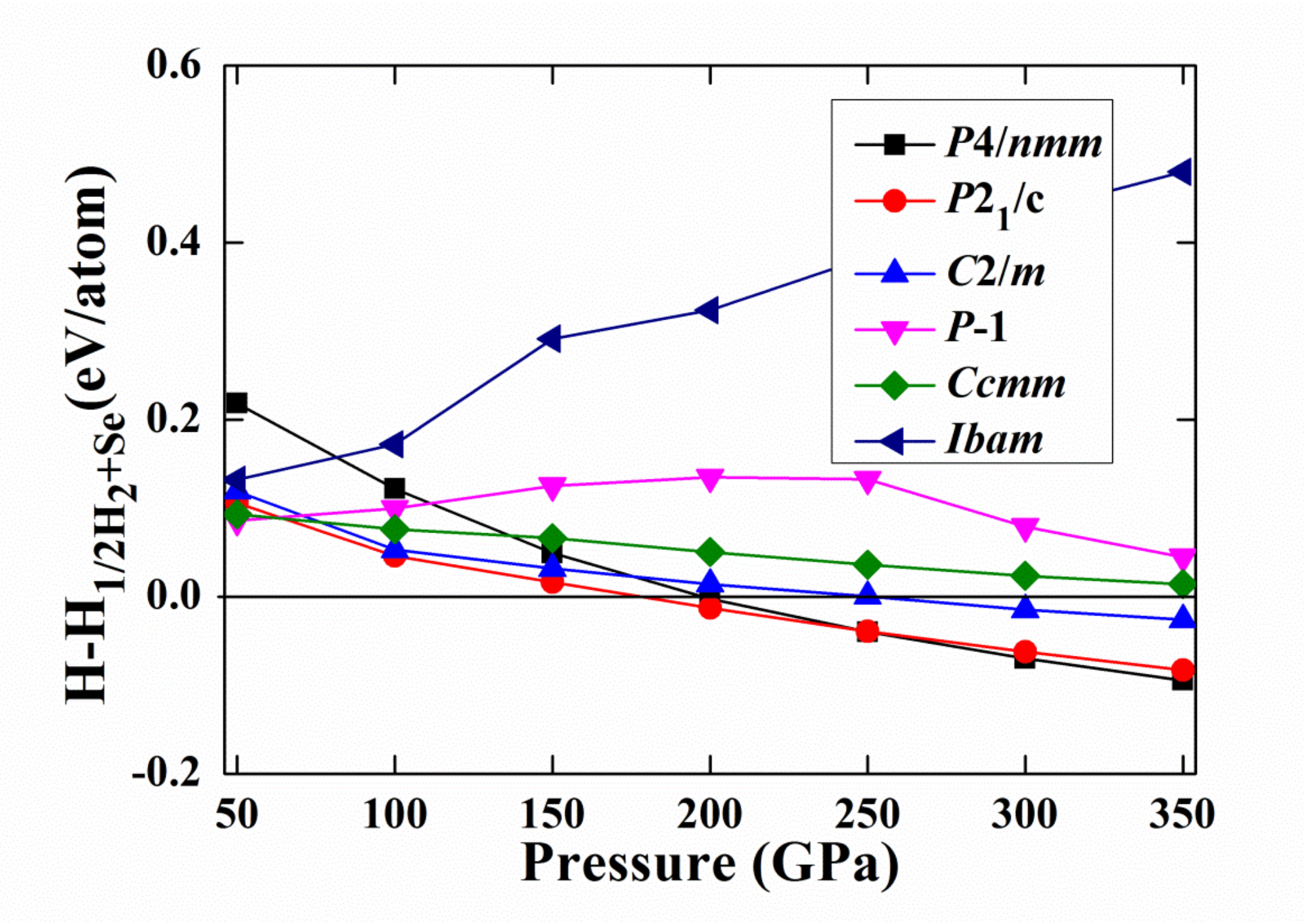

**Figure S6.** Calculated enthalpies per atom of various structures for stoichiometry HSe as functions of pressure between 50 and 350 GPa with respect to $1/2H_2$+Se.

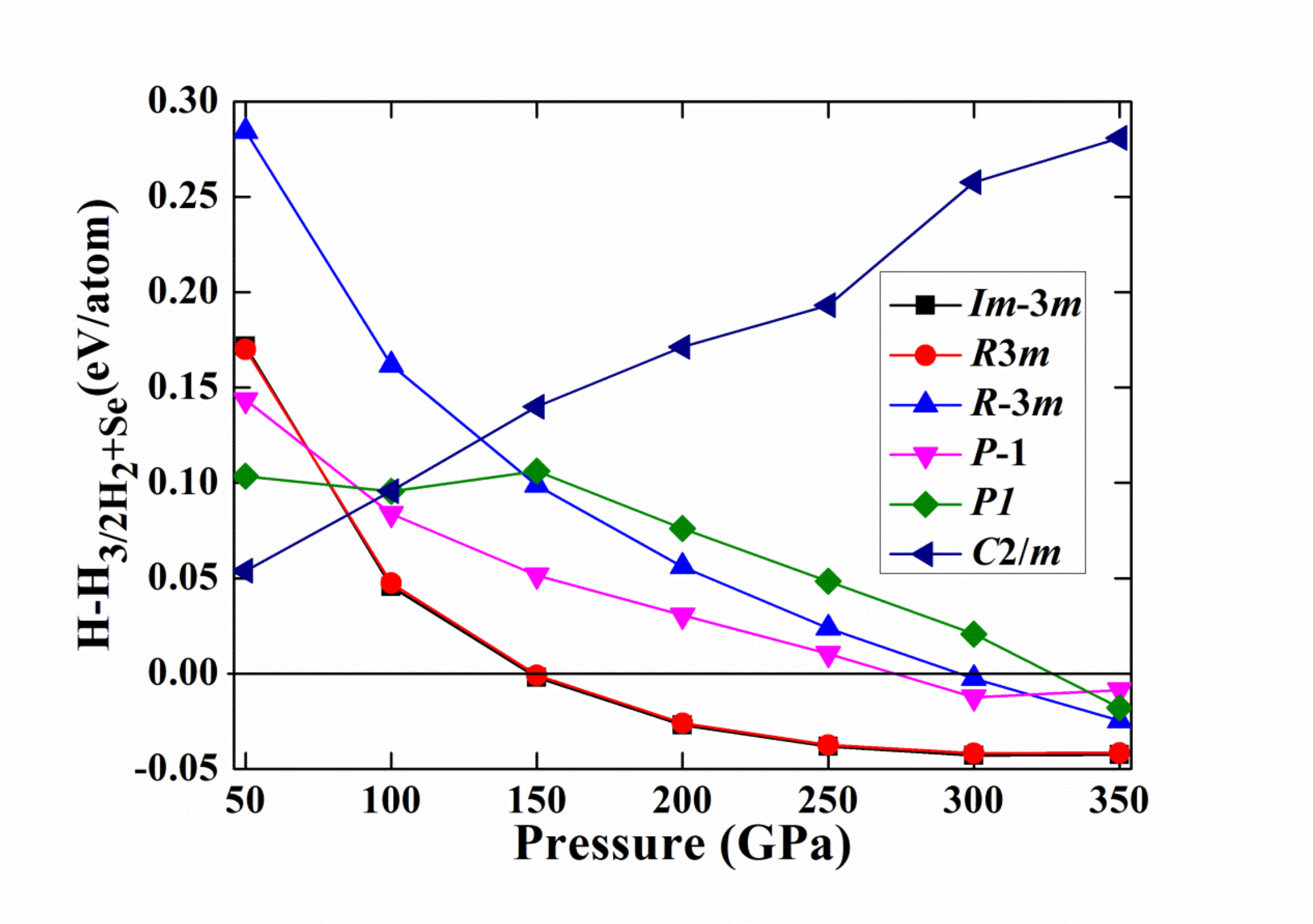

**Figure S7.** Calculated enthalpies per atom of various structures for stoichiometry $H_3Se$ as functions of pressure between 50 and 350 GPa with respect to $3/2H_2$+Se.

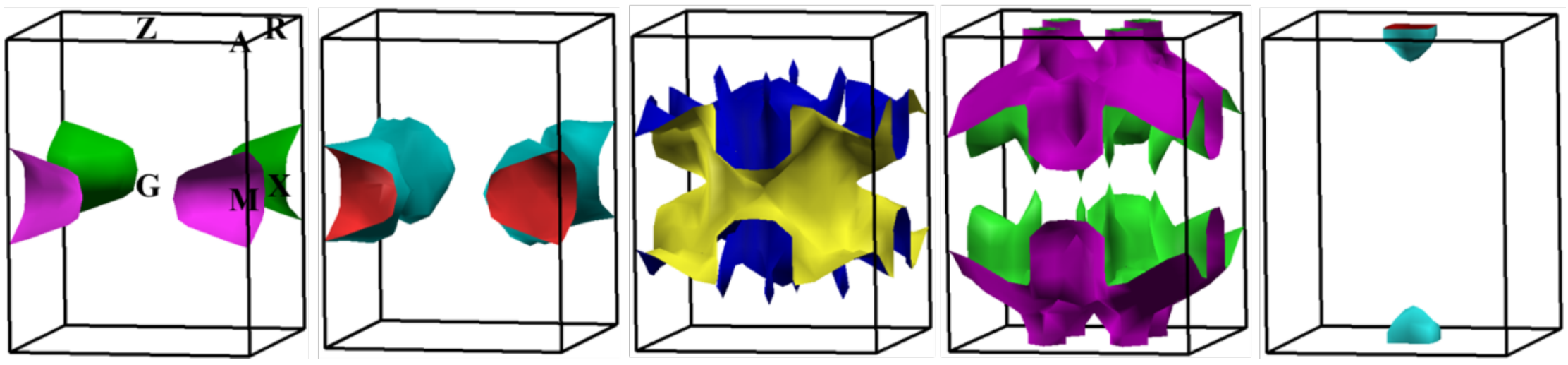

**Figure S8.** Fermi surfaces for HSe in the *P*4/*nmm* structure at 250 GPa.

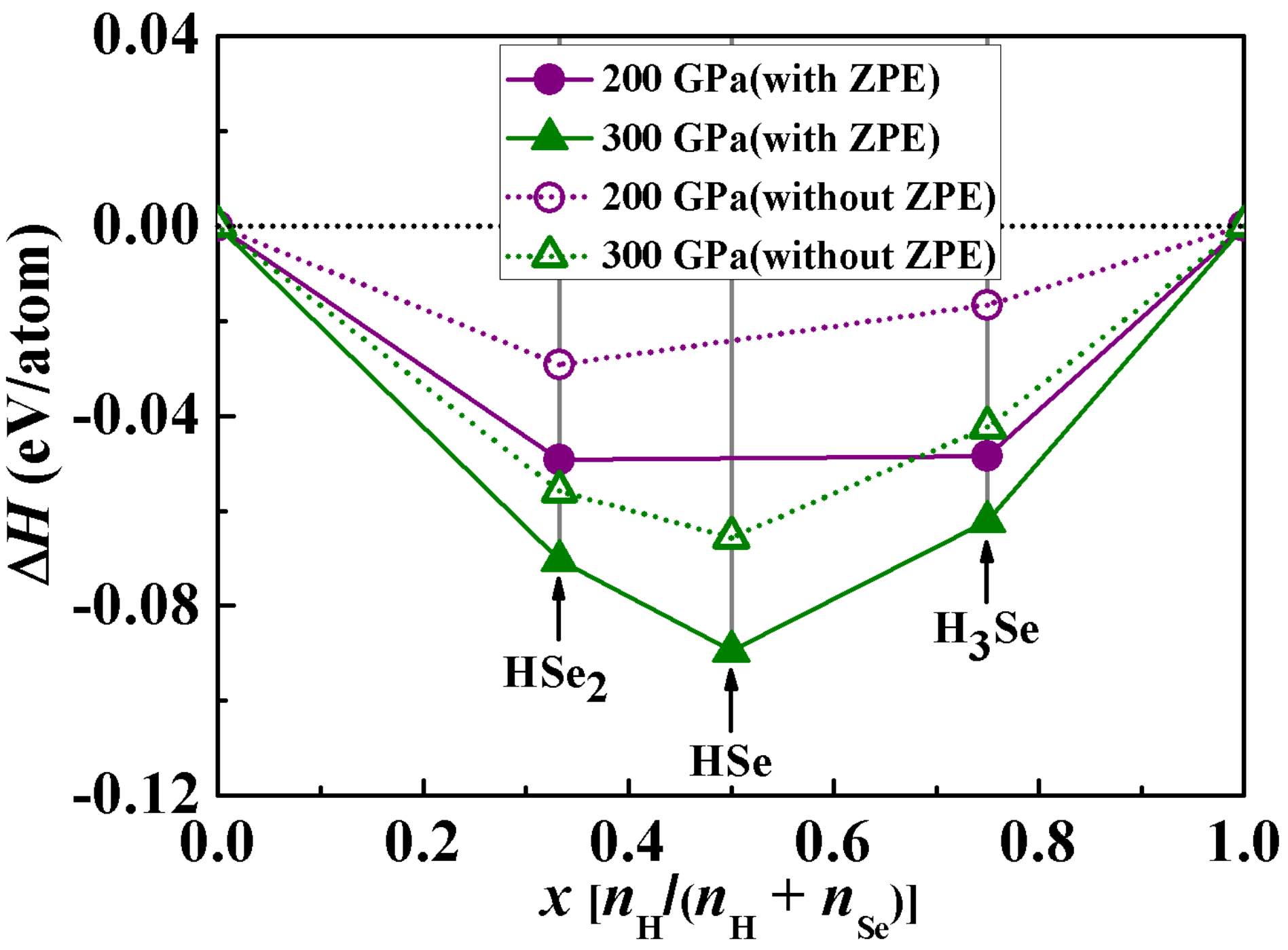

**Figure S9.** Relative formation enthalpies of the predicted stable stoichiometries, *e.g.* $HSe_2$, HSe and $H_3Se$ at 200 and 300 GPa with (filled symbols and solid lines) and without (open symbols and dash lines) the inclusion of zero-point energy contribution. One clearly sees that the introduction of zero-point energy does not change energetic stabilities of these stable stoichiometries.

# Supplemental Tables

**Table S1.** Detailed structural information of the predicted stable H-Se compounds at selected pressures.

| Phases | Pressure (GPa) | Lattice parameters (Å, °) | Atomic coordinates (fractional) | | | |
|---|---|---|---|---|---|---|
| **$HSe_2$-*C2/m*** | 300 | $a = 7.580$ | H(4i) | 0.376 | 0.000 | 0.288 |
| | | $b = 3.222$ | Se(4i) | 0.105 | 0.000 | 0.842 |
| | | c = 3.858 | Se(4i) | 0.355 | 0.000 | 0.672 |
| | | $a = \gamma = 90.000$ | | | | |
| | | $\beta = 120.320$ | | | | |
| **HSe-*P4/nmm*** | 300 | $a = b =3.014$ | H(2b) | 0.500 | 0.500 | 0.500 |
| | | c = 2.373 | Se(2c) | 0.500 | 0.000 | 0.817 |
| | | $\alpha = \beta = \gamma = 90.000$ | | | | |
| **$H_3Se$-*Im*-3*m*** | 300 | $a = b = c = 3.025$ | H(6b) | 0.500 | 0.000 | 0.500 |
| | | $\alpha = \beta = \gamma = 90.000$ | Se(2a) | 0.500 | 0.500 | 0.500 |
| **$HSe_2$-*C2/m*** | 200 | $a = 7.897$ | H(4i) | 0.376 | 0.000 | 0.288 |
| | | $b = 3.310$ | Se(4i) | 0.105 | 0.000 | 0.842 |
| | | $c = 4.048$ | Se(4i) | 0.355 | 0.000 | 0.672 |
| | | $\alpha = \gamma =90.000$ | | | | |
| | | $\beta = 120.341$ | | | | |
| **$H_3Se$-*Im*-3*m*** | 200 | $a = b = c = 3.148$ | H(6b) | 0.500 | 0.000 | 0.500 |
| | | $\alpha = \beta = \gamma = 90.000$ | Se(2a) | 0.500 | 0.500 | 0.500 |
| **$H_2Se$-$P3_121$** | 0 | $a = b =5.873$ | H(6c) | 0.900 | 0.450 | 0.030 |
| | | $c = 6.871$ | Se(3b) | 1.320 | 0.320 | 0.500 |
| | | $\alpha = \beta = 90.000$ | | | | |
| | | $\gamma = 120.000$ | | | | |

**Table S2.** Detailed structural information of the predicted metastable H-Se compounds at selected pressures.

| Phases | Pressure (GPa) | Lattice parameters (Å, °) | Atomic coordinates (fractional) | | | |
|---|---|---|---|---|---|---|
| **$H_2Se$-*C2/m*** | 300 | $a = 6.983$ | H(4i) | 0.064 | 0.000 | 0.596 |
| | | $b = 2.937$ | H(4i) | 0.094 | 0.000 | 0.965 |
| | | $c = 2.389$ | Se(4i) | 0.144 | 0.500 | 0.261 |
| | | $\alpha = \gamma = 90.000$ | | | | |
| | | $\beta = 95.442$ | | | | |
| **$HSe_2$-*C2/c*** | 300 | $a = 8.253$ | H(4e) | 0.000 | 0.523 | 0.250 |
| | | $b = 2.363$ | Se(8f) | 0.132 | 0.184 | 0.089 |
| | | $c = 4.286$ | | | | |
| | | $\alpha = \gamma = 90.000$ | | | | |
| | | $\beta = 104.475$ | | | | |
| **$HSe_2$-*Pmna*** | 300 | $a = 5.899$ | H(4g) | 0.250 | 0.200 | 0.250 |
| | | $b = 2.342$ | Se(4f) | 0.312 | 0.500 | 0.000 |
| | | $c = 5.869$ | Se(4h) | 0.000 | 0.890 | 0.187 |
| | | $\alpha = \beta = \gamma = 90.000$ | | | | |
| **$H_3Se$-*C2/m*** | 50 | $a = 6.476$ | H(2b) | 0.000 | 0.500 | 0.000 |
| | | $b = 4.875$ | H(2c) | 0.500 | 0.500 | 0.500 |
| | | $c = 4.503$ | H(8j) | 0.220 | 0.194 | 0.420 |
| | | $\alpha = \gamma = 90.000$ | Se(4i) | 0.259 | 0.000 | 0.033 |
| | | $\beta = 136.386$ | | | | |
| **$H_3Se$-*R3m*** | 200 | $a = b = 4.451$ | H(9b) | 0.500 | 0.500 | 0.298 |
| | | $c = 2.728$ | Se(3a) | 0.000 | 0.000 | 0.298 |
| | | $\alpha = \beta = 90.000$ | | | | |
| | | $\gamma = 120.000$ | | | | |
| **$H_3Se$-*P*-1** | 300 | $a = 2.588$ | H(2i) | 0.335 | 0.010 | 0.505 |
| | | $b = 2.565$ | H(2i) | 0.277 | 0.623 | 0.371 |
| | | $c = 4.279$ | H(2i) | 0.422 | 0.948 | 0.130 |
| | | $\alpha = 91.920$ | Se(2i) | 0.865 | 0.349 | 0.217 |
| | | $\beta = 94.937$ | | | | |
| | | $\gamma = 73.627$ | | | | |
| **HSe-*$P2_1/c$*** | 300 | $a = 4.081$ | H(1a) | 0.924 | 0.412 | 0.193 |
| | | $b = 4.456$ | H(1a) | 0.075 | 0.588 | 0.806 |
| | | $c = 2.563$ | H(1a) | 0.575 | 0.912 | 0.306 |
| | | $\alpha = \gamma = 90.000$ | H(1a) | 0.424 | 0.088 | 0.693 |
| | | $\beta = 78.453$ | Se(1a) | 0.697 | 0.640 | 0.631 |
| | | | Se(1a) | 0.302 | 0.359 | 0.368 |
| | | | Se(1a) | 0.802 | 0.140 | 0.868 |
| | | | Se(1a) | 0.197 | 0.859 | 0.131 |

**Table S3.** Calculated electron-phonon coupling parameter λ, logarithmic average frequency $\omega_{\log}$, electronic density of states at the Fermi level *N* (*Ef*) and critical temperature $T_c$ of predicted H-Se compounds and $H_3S$.

| **Phases** | **Pressure (GPa)** | **λ** | **$\omega_{\log}$ (K)** | **$N(E_f)$ states/spin/Ry/cell** | **$T_c$ (K)** |
|---|---|---|---|---|---|
| **$HSe_2$-*C2/m*** | 300 | 0.45 | 647 | 8.09 | 5 |
| **HSe-*$P2_1/c$*** | 300 | 0.65 | 832 | 9.55 | 23 |
| **HSe-*P4/nmm*** | 250 | 0.81 | 813 | 5.62 | 39 |
| | 300 | 0.80 | 885 | 5.75 | 42 |
| **$H_3Se$-*Im-3m*** | 200 | 1.09 | 1477 | 3.19 | 116 |
| | 250 | 1.04 | 1498 | 3.13 | 111 |
| | 300 | 1.04 | 1492 | 3.09 | 110 |
| **$H_3S$-*Im-3m*** | 200 | 1.61 | 1484 | 3.27 | 171 |
| | 250 | 1.33 | 1708 | 3.25 | 172 |
| | 300 | 1.20 | 1774 | 3.23 | 160 |